\documentclass[aps,twocolumn]{revtex4}
\usepackage{amssymb}
\usepackage{amsmath}
\usepackage{epsfig}
\begin{document}
\newcommand{\sn}{\,\rm sn}
\newcommand{\cn}{\,\rm cn}
\newcommand{\dn}{\,\rm dn}
\newcommand{\cd}{\,\rm cd}
\title{Nonlinear states and nonlinear tunneling \\
in a potential well }
\author{ J\'er\^ome LEON}
 \affiliation{ Physique Math\'ematique et Th\'eorique, CNRS-UMR5825\\
Universit\'e Montpellier 2, 34095 MONTPELLIER  (France)}

\begin{abstract} A nonlinear Schr\"odinger model in a square well and managed
nonlinearity is shown to possess nonlinear states as continuous extensions of
the linear levels. The solutions are remarkably stable up to a threshold
amplitude where a soliton is emitted and propagates outside the well. The
analytic expression of the threshold is given in terms of the well size for
each level. This process of {\em nonlinear tunneling} results from an
instability of the evanescent wave inside the walls and can find experimental
realization in a proposed nonlinear fiber Bragg gratings resonator.
\end{abstract}

\maketitle

\paragraph*{Introduction.}

We consider the nonlinear Schr\"odinger equation (NLS) for the {\em wave
function}  $\psi(x,t)$ trapped in a {\em square well}, in reduced units,
according to the following model
\begin{align}
&|x|> L\ :\ &&i\psi_t+\psi_{xx}+|\psi|^2\psi=V\psi, \notag\\
&|x|< L\ :\ &&i\psi_t+\psi_{xx}-|\psi|^2\psi=0.\label{model}
\end{align}
The  positive constant $V$ is the well height and $2L$ its width.
The nonlinearity is managed such as to be {\em repulsive}  inside the well 
and {\em attractive}  outside.

The nonlinear states for NLS with a definite sign of the nonlinearity has been
derived in \cite{carr} with interesting solutions having no counterparts in the
linear limit. Using nonlinearity management as in \eqref{model} actually allows
to obtain a system whose nonlinear states solutions possess remarkable
properties: i) they uniformly tend to the linear eigenstates in the small
amplitude limit, ii) they are stable solutions for amplitudes below an explicit
threshold, iii) above the threshold an instability generates gap
solitons, emitted outside the well, hence realizing a classical nonlinear
tunneling process.

\begin{figure}[ht] \centerline {\epsfig{file=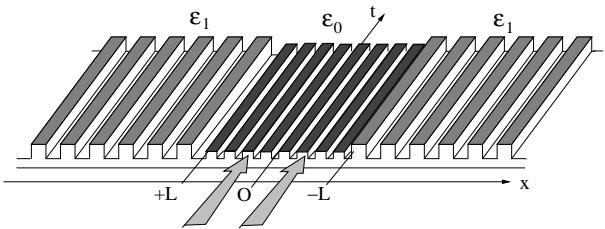,height=3cm,width=8cm}}
\caption {Example of a physical situation where the model \eqref{model} applies
where $t$ is the propagation direction and  $V$ is proportionnal to the 
dielectric constant difference $\epsilon_0-\epsilon_1$.}
\label{fig:grating}\end{figure}
This problem has promising application to nonlinear optics where it applies to
the  propagation in the direction $t$, and tranverse modulation along $x$, of
the envelope of a laser beam in a fiber Bragg grating (in Kerr regime) 
\cite{kivsh-agra} when different types of fibers are used: self-focusing
nonlinearity with dielectric constant $\epsilon_1$ outside $[-L,+L]$,
self-defocusing with $\epsilon_0>\epsilon_1$ inside. The periodic medium
extending in $|x|>L$ acts as a Bragg mirror for the tranverse modulation and
the resulting {\em Bragg grating resonator} is sketched on figure
\ref{fig:grating}. The arrows show the injected radiation.

The domain of spatial solitons in fiber Bragg gratings is rich of recent
results, see e.g. \cite{kivshar,khomeriki}, with nice experimental evidence of
soliton formation \cite{mandelik} and recent experiments which demonstrated the
existence of {\em discrete stationnary} gap solitons \cite{expe}. On the
theoretical side, an interesting proposal of \cite{khomeriki} is to generate 
{\em discrete gap solitons}  by boundary driving a Bragg grating  {\em above
the cut-off}. The aproach uses the theory of nonlinear supratransmission
\cite{nst} extended to the case when the forbidden band results from the
{\em discrete} nature of the medium.

\begin{figure}[ht] \centerline
{\epsfig{file=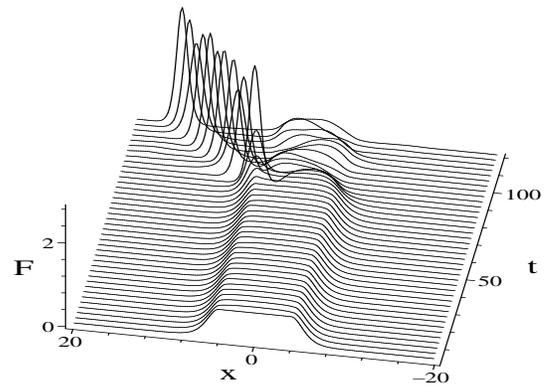,height=5cm,width=7cm}} \caption {Profile of
$F=|\psi|^2$ for the fundamental 0-mode  initial datum at threshold amplitude
in the case $V=1$ and $L=4$.} \label{fig:tunneling}\end{figure}

In the quasi-continous case (large number of fibers and weak coupling), the
device of fig.\ref{fig:grating} will be shown to be a means to generate gap
solitons by continuous wave (cw) input radiation at a  flux intensity given
explicitely in terms of the dimensions  of the potential well depth $V$ (or the
dielectric constant varaition $\epsilon_0-\epsilon_1$).  As an illustration,
fig.\ref{fig:tunneling} shows the evolution of the fundamental mode at the
threshold for which the input envelope is indeed cw-like and experiences an
instability generating the gap soliton.
The threshold vaues are explicitely calculated for the entire set of nonlinear
states. Finally we demonstrate that the linear small amplitude limit maps
continuously the nonlinear states to the linear ones.

We shall need to refer to the linear eigenstates
that read in the odd case
\begin{align}
&|x|\le L\ :\ &&\psi=A\,e^{-i\omega t}\sin(k x)\ ,\notag\\
&|x|\ge L\ :\ &&\psi=A\,e^{-i\omega t}\sin(k L)e^{-\kappa(|x|-L)}\ ,
\notag\\
& k^2=V\sin(kL)\ ,&& \tan(kL)<0\ .\label{odd-lin}
\end{align}
and in the even case
\begin{align}
&|x|\le L\ :\  &&\psi=A\,,e^{-i\omega t}\cos(k x)\ ,\notag\\
&|x|\ge L\ :\  &&\psi=A\,e^{-i\omega t}\cos(k L)e^{-\kappa(|x|-L)}\ ,
\notag\\
& k^2=V\cos(kL)\ ,&& \tan(kL)>0\ ,\label{even-lin}
\end{align}
for $\omega=k^2<V$ and  $\kappa^2=V-\omega$, and for arbitrary amplitude $A$.

\paragraph*{Nonlinear states.}

As learned from \cite{carr}, the main tool to derive the nonlinear states  is
to connect a periodic solution inside the well to a static one-soliton  tail
outside.  We make use of the fundamental solutions of NLS in terms of Jacobi
elliptic functions for both symmetric (even) and antisymmetric (odd) cases 
\cite{jacobi}.

First, outside the potential well we have the common 
soliton tails given by (still with $\kappa^2=V-\omega$ and $\omega<V$)
\begin{equation}\label{tails}
|x|\ge L,\quad 
\psi=\frac {\kappa\sqrt2\,e^{-i\omega t}}{\cosh [\kappa(|x|-L+d)]}\ ,
\end{equation}
which replaces the two evanescent waves of the linear case. Note that  the
function $(-\psi)$ is also solution, a property which must be used to
conveniently connect the solution inside to the outside tails.

Inside the well we use the basic solution 
\begin{equation}\label{basic}
|x|\le  L,\quad \psi=\mu A\, e^ {-i\omega t}\sn (A' x+x_0,\mu) .
\end{equation}
Here and in the following the {\em amplitude} $A$ is real-valued and positive
and we define the new constant $A'$ by $A=A'\sqrt2$. By convenient choices
of the constant $x_0$ we shall obtain both the odd and even solutions in the
well.

This function $\psi$ is a solution of \eqref{model} if (necessary condition)
\begin{equation}\label{omega-mu}
\omega^2=\frac12 A^2(1+\mu^2),\end{equation}
which links the frequency $\omega$ to the modulus $\mu$ of the elliptic
functions. The requirement $\omega<V$ for a bound state implies that the
modulus $\mu$ cannot exceed a maximum value $\mu_m$
\begin{equation}\label{mu-max}
\mu\le(2V/A^2-1)^{1/2}=\mu_m.\end{equation}

We discovered that it is necessary to work with expressions of the solution 
for moduli $\mu>1$, performed by means of the identity 
$\mu\sn(u,\mu)=\sn(\mu u,1/\mu)$. The solutions for $\mu<1$ can be easily 
written down but either they do not contribute to the spectrum or they have
their exact conterpart with $\mu>1$.

\paragraph*{Odd solutions.} 

In the range $\mu\in[1,\mu_m]$ we define the following solution inside the well 
\begin{equation}\label{odd}
|x|\le L,\quad \psi=A\, e^ {-i\omega t}\sn (\mu A' x,\frac 1\mu)\ .
\end{equation}
In order to obtain sufficient conditions for the set \eqref{odd} 
\eqref{tails} to be a solution of \eqref{model}, we need to ensure
continuity of the solution and its derivative in $x=\pm L$. This provides
the admissible discrete set of moduli $\mu$ which, after some algebra,
must solve
\begin{align}\label{cont}
\mu^2 & \cn^2(b,\frac 1\mu) \dn^2(b,\frac 1\mu)=\notag\\
&\sn^2(b,\frac 1\mu)
\left[\frac{2V}{A^2}-(1+\mu^2)-\sn^2(b,\frac 1\mu)\right],\end{align}
where  the function $b$ is defined as
\begin{equation}\label{b-odd}b=\mu A'L.\end{equation}
Then the shift  $d$ of the tail position for each solution $\mu$ of the above
equation, is given by
\begin{equation}\label{d}
A|\sn (b,\frac 1\mu)|=\frac{\kappa\sqrt2}{\cosh (\kappa d)}.
\end{equation}

As for the linear case we must add a consistency condition which
ensures the continuity of the derivatives. It is just a matter of careful
reading of all possible cases to obtain the condition
\begin{equation}\label{consistency}
\sn (b,\frac 1\mu)\cn (b,\frac 1\mu)\dn (b,\frac 1\mu)<0,\end{equation}
for the solution $\mu$ of \eqref{cont} to be acceptable.

\paragraph*{Even solutions.} 
Still with $\mu\in[1,\mu_m]$ we define 
\begin{equation}\label{even}
|x|<L,\quad\psi=A\, e^ {-i\omega t}\sn (\mu A' x+K(\frac1\mu),\frac 1\mu)\ .
\end{equation}
obtained from \eqref{basic} by the translation  $x_0=K(1/\mu)/\mu$ where
$K$ is the complete elliptic integral of the first kind.
In that case the continuity conditions give the admissible
set of moduli as solutions of \eqref{cont} with now
\begin{equation}\label{b-even}
b=\mu A' L+K(\frac1\mu)
\end{equation}
and the consequent definition \eqref{d} of $d$. The consistency condition
\eqref{consistency} still holds here with the above function $b$.

Note that in both expressions \eqref{odd} and \eqref{even}, the parameter $A$
does represent the maximum value of the amplitude of the solution, reached at
$x=0$ for the even states  \eqref{even} and at $x=K(1/\mu)$ for the odd states
\eqref{odd}.

\begin{figure}[ht]
\centerline {\epsfig{file=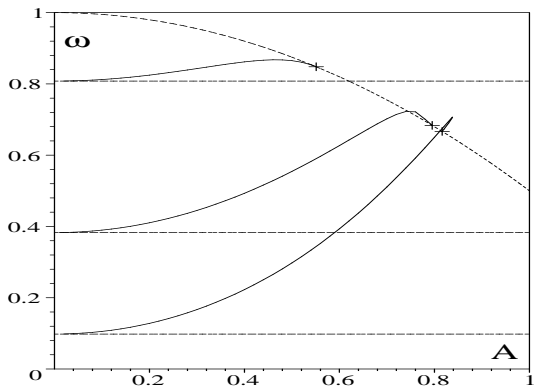,height=5cm,width=7cm}}
\caption {Dependence of the eigenvalues $\omega$ in terms of the 
amplitude for $V=1$ and $L=4$. Crosses indicate the values of amplitude 
threshold and the dashed curve is the prediction of global threshold 
\eqref{threshold}.}\label{fig:eigenmodes}\end{figure}

\paragraph*{Example.}
The figure \ref{fig:eigenmodes} shows the solutions of the equations
\eqref{cont} for $V=1$ and $L=4$  in the odd and even cases, for which the
associated linear problem possess 3 eigenstates (dashed lines). The 3 curves 
show the dependency of the eigenvalue of each nonlinear extension of the linear
levels in terms of the amplitude $A$. These eigenvalues are given by the
expression \eqref{omega-mu} for each solution $\mu$ of  \eqref{cont}, in both
cases \eqref{b-odd} and \eqref{b-even}, equations that are solved numerically
(like in the linear case).

The next figure \ref{fig:3modes} displays the three states for $V=1$ in the
linear and nonlinear case corresponding to particular choices of amplitudes and
frequencies as indicated on the graphs. These are the plots of the solutions
\eqref{odd}  and \eqref{even}, compared to the linear eigenfunctions 
\eqref{odd-lin} and \eqref{even-lin}.
\begin{figure}[ht]
\centerline {
\begin{tabular}{cc}
LINEAR & NONLINEAR\\
 {\epsfig{file=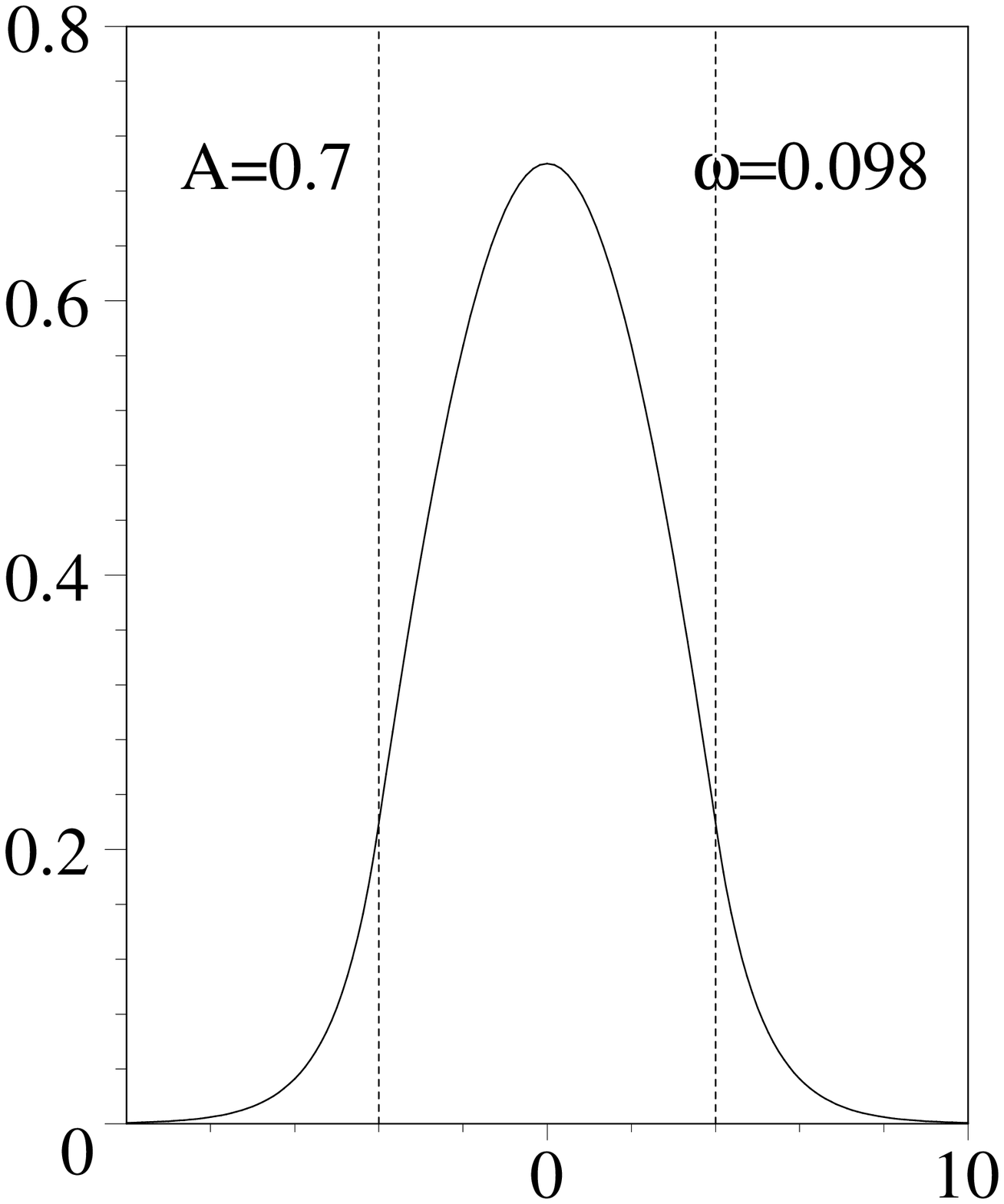,height=2.5cm,width=4cm}} &
 {\epsfig{file=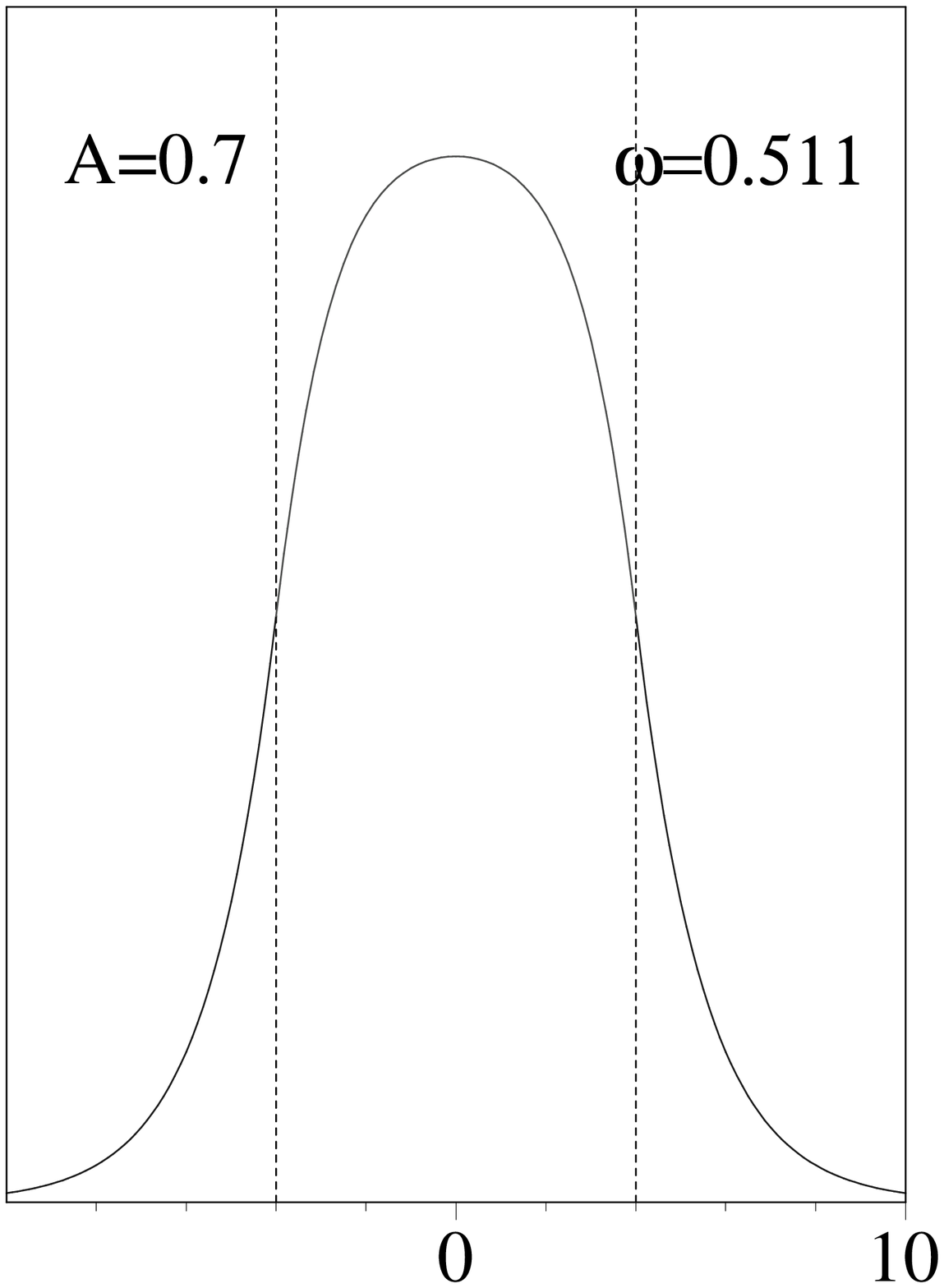,height=2.5cm,width=3.6cm}}\\
 {\epsfig{file=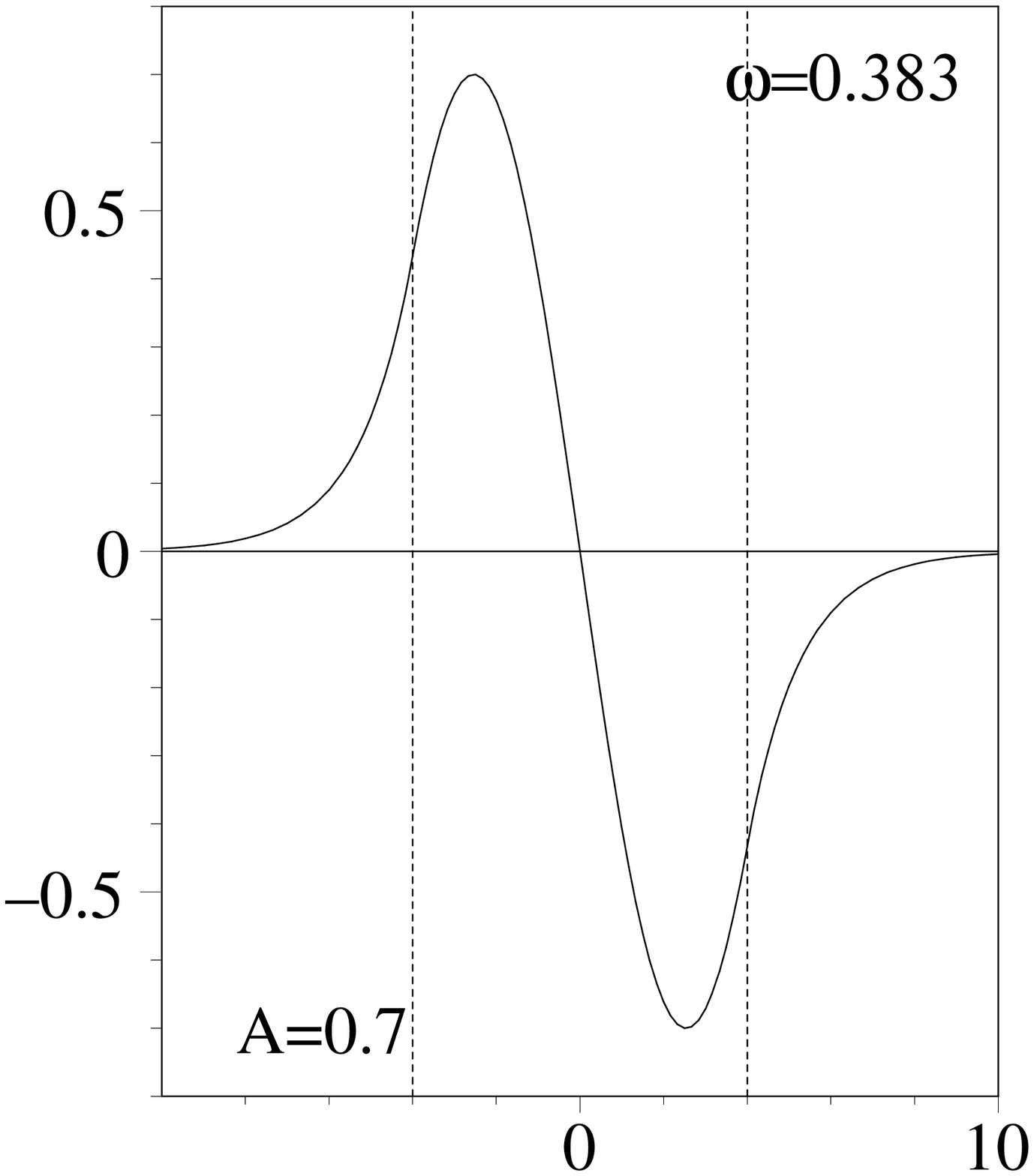,height=2.5cm,width=4.2cm}} &
 {\epsfig{file=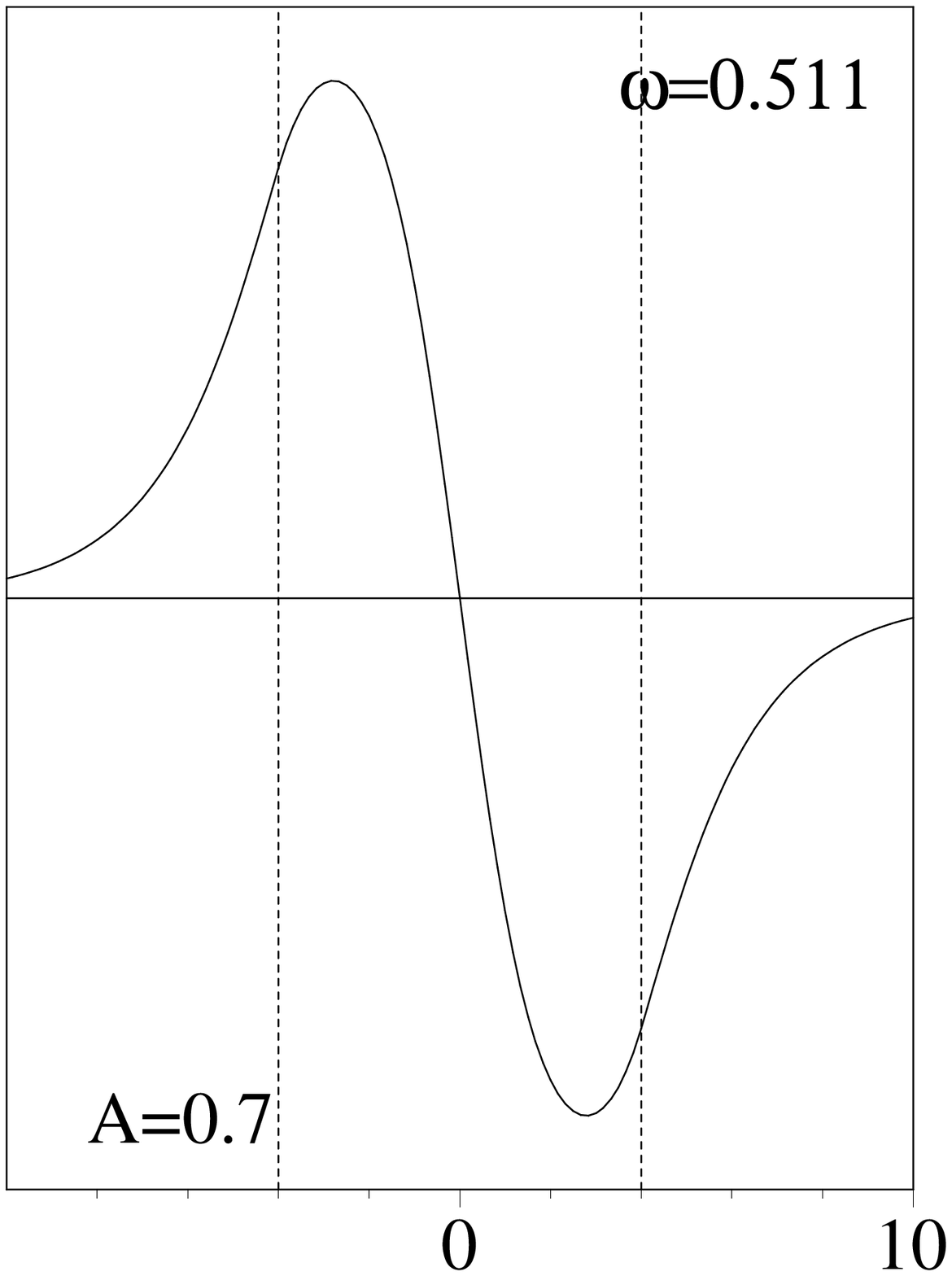,height=2.5cm,width=3.6cm}}\\
 {\epsfig{file=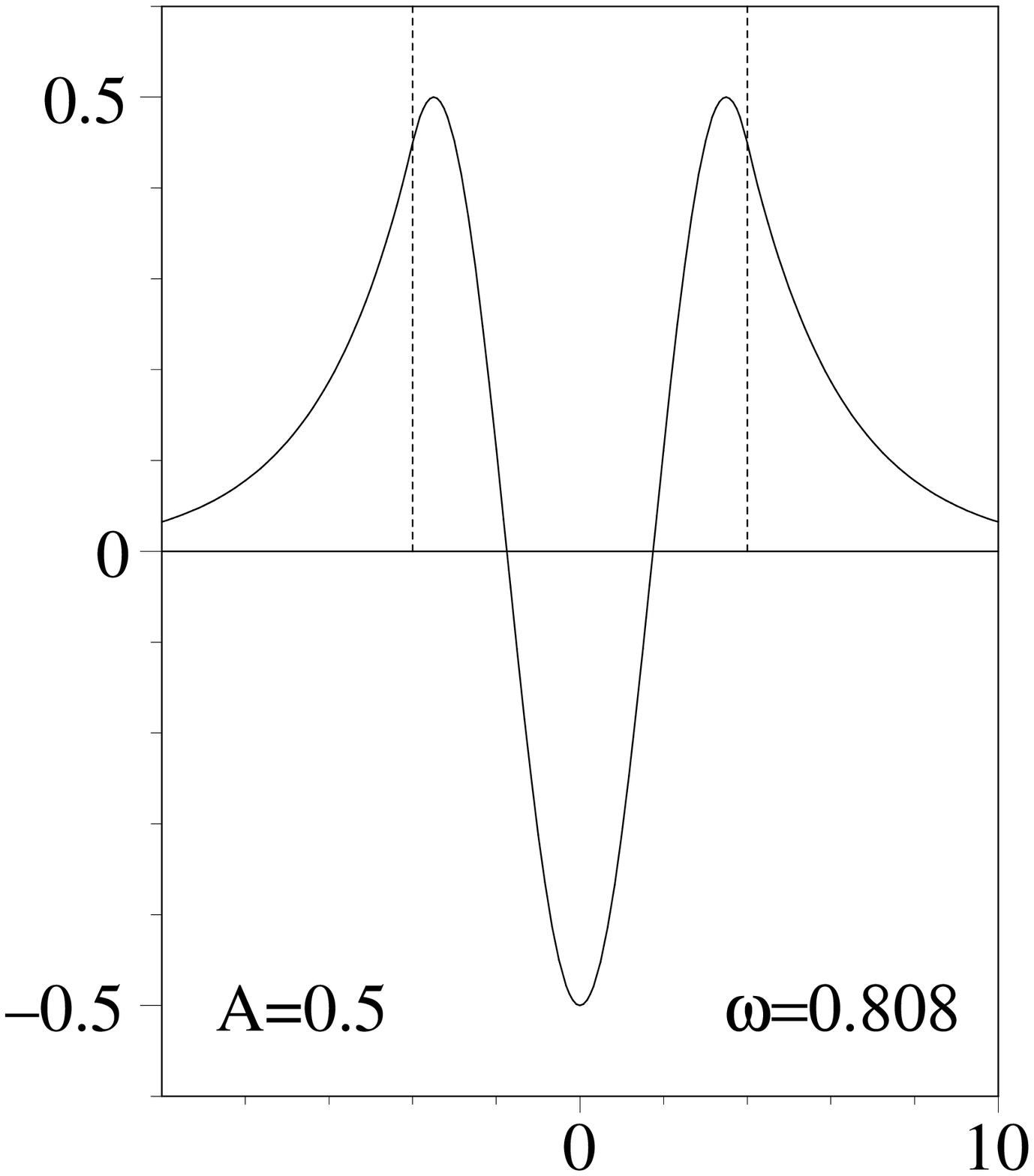,height=2.5cm,width=4.2cm}}&
 {\epsfig{file=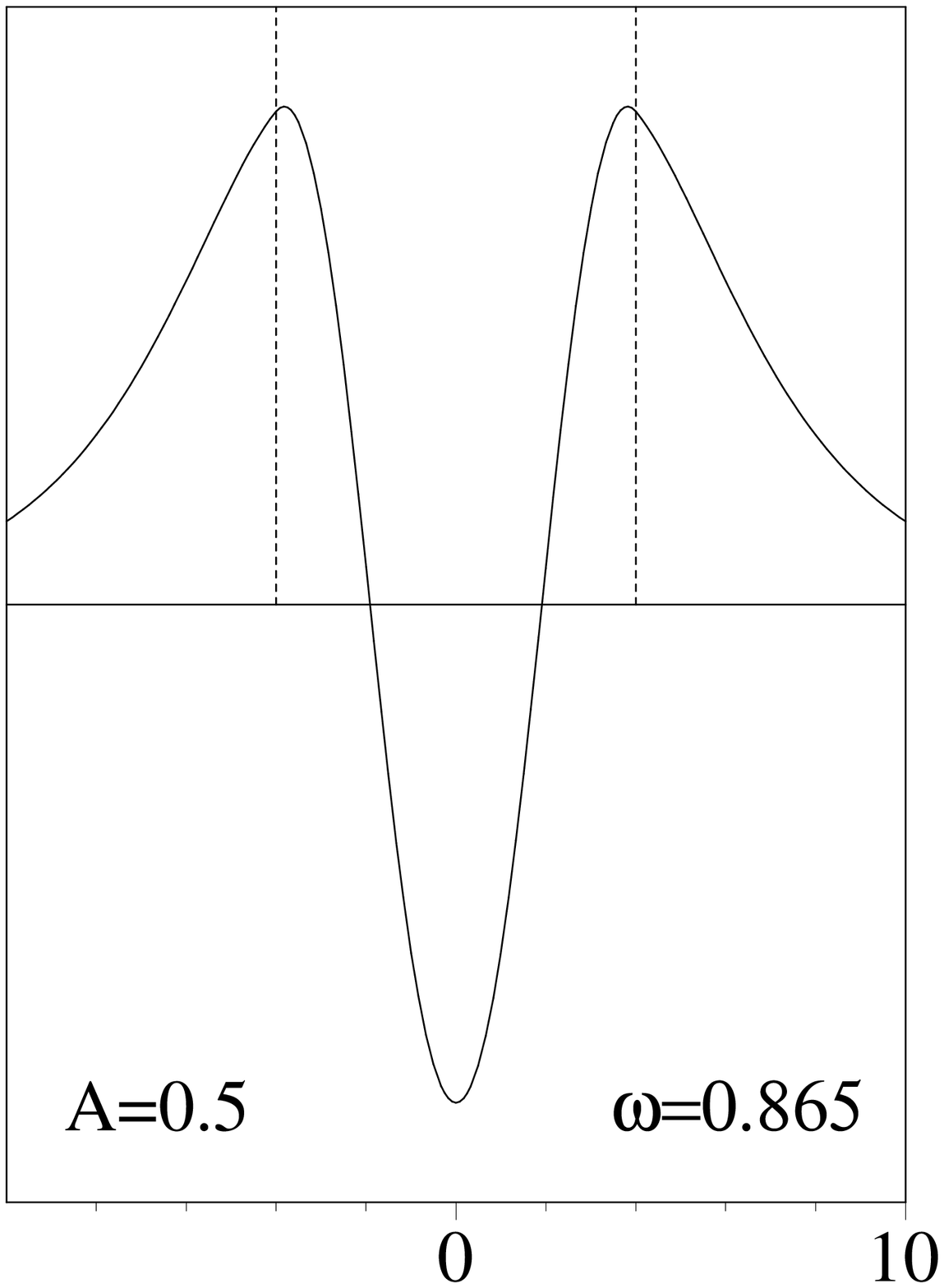,height=2.5cm,width=3.6cm}}\end{tabular}}
 \caption {Examples of linear and nonlinear states, for $V=1$,
with amplitudes and frequencies as indicated. The same vertical scale is used 
for linear and nonlinear figures.}\label{fig:3modes}\end{figure}

There are two fundamental properties of this {\em ``nonlinear spectrum''} that
we demonstrate hereafter. It is first the evident natural property to reach the
linear  spectrum for  vanishing amplitudes.  This explains in particular why we
have the same number of states as the linear levels. Second all three curves are
seen to stop at some threshold amplitude which can be exactly computed, as
shown by the dashed curve on figure \ref{fig:eigenmodes} and the crosses which
are the predicted thresholds.

\paragraph*{Linear limit.}

To demonstrate in general that the nonlinear spectrum goes to the linear one in
the limit $A\to0$, it is crucial to evaluate the limits keeping the product
$A\mu$ finite. Actually defining  
\begin{equation}
k=\frac A{\sqrt{2}}\mu,\end{equation}  
we can easily obtain ($A=A'\sqrt{2}$)
\begin{equation}\label{limit} A\sn (\mu A'x,1/\mu)
\underset{A\to0}{\sim} A\sin(kx), \end{equation}  
which proves that the nonlinear odd states \eqref{odd} tend to the linear
ones in the limit $A\to0$. 
The same property holds naturally for the nonlinear even states \eqref{even} by
using $\sn(u+K(\nu),\nu)=\cd(u,\nu)$  and $\cd(u,0)=\cos(u)$ together with
\eqref{limit} above.

These results show in particular that the {\em ``nonlinear wave number''} is
the quantity $k=A'\mu$ and thus that the relation \eqref{omega-mu} must be
read  $\omega^2=k^2+A^2/2$ that tends to the linear dispersion relation
$\omega^2=k^2$ in the small amplitude limit.

\paragraph*{Nonlinear tunneling.}

More interesting is the existence of a threshold amplitude (or population
threshold) beyond which gap solitons are emitted outside the well hence
realizing a classical nonlinear tunneling. This is the result of an
instability, as described in \cite{jl-pla}, which takes place as soon as the
amplitude $A$ is such that the soliton tails \eqref{tails} reaches its
maximum value in $|x|=L$, i.e. $d=0$.

In order to derive the analytic expression of the thresholds postions
$\{A_s,\omega_s\}$ for each branch $\omega(A)$, we express the continuity
conditions in $|x|=L$ in the case $d=0$. The threshold is thus obtained as a
particular solution of the equation for the state \eqref{cont}  for which both
sides vanish altogether (successively for the odd and even states), namely
\begin{align}
& \cn^2(b,\frac 1\mu) \dn^2(b,\frac 1\mu)=0,\label{seuil-1}\\
&\frac{2V}{A^2}-(1+\mu^2)-\sn^2(b,\frac 1\mu)=0.\label{seuil-2}
\end{align}
where $b$ is given by \eqref{b-odd} for the odd solutions and by \eqref{b-even}
for the even ones.

It is then a matter of algebraic manipulations to demonstrate that the solutions
$\{A_s,\ \mu_s\}$ of the above equations globally satisfy
\begin{equation}\label{threshold}
A_s^2=\frac{2V}{2+\mu_s^2}=2(V-\omega_s),\end{equation}
where $\mu_s$ is obtained by solving for the odd case
\begin{equation}\label{seuil-odd}
\sn^2(L\mu\sqrt{\frac{V}{\mu^2+2}},\frac 1\mu)=1,\end{equation} 
and for the even case
\begin{equation}\label{seuil-even}
\sn^2(L\mu\sqrt{\frac{V}{\mu^2+2}}+K(\frac 1\mu),\frac 1\mu)=1.\end{equation} 
Note that the solutions $\{A_s,\ \mu_s\}$ of these equation must obey
requirement \eqref{mu-max} which is checked a posteriori.

The dashed curve of fig. \ref{fig:eigenmodes} is the plot of \eqref{threshold}
for $V=1$ and $L=4$, and the crosses are obtained by solving \eqref{seuil-odd} 
and \eqref{seuil-even} numerically.

The above pocedure does not furnishes the threshold corresponding to the
fundamental level. Indeed it misses the particular solution 
\begin{equation}\label{seuil-fond}
\mu_s=1,\quad A_s^2=\frac23V,\end{equation}
which is effectively a solution of \eqref{seuil-1}\eqref{seuil-2} from the
property
\begin{equation}
\lim_{\mu\to1}\left\{\sn(a+K(\frac 1\mu),\frac 1\mu)\right\}=1,
\end{equation}
for any real-valued $a$. In that case the {\em ``dispersion relation''}
\eqref{omega-mu} provides $\omega_s=A_s^2$. Note that the threshold
amplitude $A_s$ does not depend on the width $2L$ of the well.

This solution is particularly interesting in view of applications to the fiber
guide grating of fig.\ref{fig:grating}. Indeed in that case the corresponding
expression \eqref{even} is the constant amplitude field 
$\psi=A_se^{-i\omega_st}$ (which is trivialy a solution of \eqref{model}).
Injecting then in the medium (as indicated by the arrows on
fig.\ref{fig:grating}) a cw-laser beam of (normalized) intensity flux $A_s^2$
constant along the tranverse $x$-direction for $x\in[-L,+L]$, one would
generate a gap soliton propagating outside the well, as displayed in fig.
\ref{fig:tunneling}. Such simulation can be reproduced {\em ad libidum} for
input data constant in $[-L,+L]$ and exponentially vanishing outside. As soon
as the amplitude exceeds the threshold $A_s$, one obtains tunelling by soliton 
emission.

It is clear that the present formalism has been developped for continuous
envelopes $\psi(x,t)$, which requires a large number of fiber guides and weak
transverse coupling. An interesting extension is then the discrete situation
with the open question of the discrete states solutions.

\paragraph*{Perspectives and conclusion.}

We have so far demonstrated the existence of {\em ``nonlinear eigenstates''},
exact solutions of the NLS model \eqref{model}, that are remarkably stable
(numerical simulations did not show any deviation from the exact expressions up
to times when numerical errors become sensible, i.e. $10^3$ to $10^4$ for our
scheme) as soon as their amplitudes do not exceed a threshold explicitely
evaluated in terms of the well height for each mode. These states reproduce
exactly, in the small amplitude limit, the usual eigenstates of the
Shr\"odinger equation in a potential well.

Above the threshold amplitude, a {\em ``nonlinear tunneling''} (macroscopic,
classical) occurs by the emission of gap solitons outside the well. This
generic property suggests a {\em ``gap soliton generator''} by the device
sketched in fig.\ref{fig:grating}, more especially as the fundamental level,
close to the threshold, behaves as a cw-field.

The approach applies straightforwardly to a model where the inside of the well
would obey the linear Schr\"odinger equation. This in view of applications to a
resonator device with Bragg medium playing the role of the mirrors, a subject
under study now.

Another natural question is the properties of the nonlinear states when the
inside of the well presents an attractive (focusing) nonlinearity (as the
outside medium). In that case, similar states are defined in terms of Jacobi
$\cn$-functions \cite{carr} but we have observed that they experience
modulational instability before reaching the threshold amplitude.  In the other
case when the whole medium is defocusing, the outside tails are of cosech-type
and thus can match any amplitude of the nonlinear state, hence there is no
threshold and no gap soliton generation. Thus the management of the
nonlinearity chosen in \eqref{model}  is a means for sabilizing the nonlinear
states and ensuring a threshold for soliton generation.

We report to future studies the important question of the analytical proof of
the stability of those states, including the mathematical derivation of the
instability criterion above the threshold amplitude. Note finally that, in the
threshold regions, the  states are degenerated: two solutions coexist at the
same frequency with slightly different amplitudes as seen on
fig.\ref{fig:eigenmodes}. 

\end{document}